\documentclass[conference]{IEEEtran}
\IEEEoverridecommandlockouts
\usepackage{cite}
\usepackage{amsmath,amssymb,amsfonts}
\usepackage{algorithmic}
\usepackage{graphicx}
\usepackage{textcomp}
\usepackage{xcolor}
\usepackage{alphabeta}
\usepackage{hyperref}
\usepackage{lscape}
\def\BibTeX{{\rm B\kern-.05em{\sc i\kern-.025em b}\kern-.08em
    T\kern-.1667em\lower.7ex\hbox{E}\kern-.125emX}}

\begin{document}

\title{Is Productivity in Quantum Programming Equivalent to
Expressiveness?}

\author{\IEEEauthorblockN{Francini Corrales-Garro}
\IEEEauthorblockA{\textit{School of Software Engineering } \\
\textit{CENFOTEC University}\\
San José, Costa Rica \\
\href{mailto:fcorralesg@ucenfotec.ac.cr}{fcorralesg@ucenfotec.ac.cr} \\
ORCiD 0000-0003-3502-4128}
\and
\IEEEauthorblockN{Danny Valerio-Ramírez}
\IEEEauthorblockA{\textit{School of Software Engineering} \\
\textit{CENFOTEC University}\\
San José, Costa Rica \\
\href{mailto:dvalerior@ucenfotec.ac.cr}{dvalerior@ucenfotec.ac.cr} \\
ORCiD 0009-0002-8766-7723}
\and
\IEEEauthorblockN{Santiago N\'uñez-Corrales}
\IEEEauthorblockA{\textit{NCSA/IQUIST} \\
\textit{UIUC}\\
Urbana IL, United States \\
\href{mailto:nunezco2@illinois.edu}{nunezco2@illinois.edu} \\
ORCiD 0000-0003-4342-6223}
}

\maketitle

\begin{abstract}
The expressiveness of quantum programming languages plays a crucial role in the efficient and comprehensible representation of quantum algorithms. Unlike classical programming languages, which offer mature and well-defined abstraction mechanisms, quantum languages must integrate cognitively challenging concepts such as superposition, interference and entanglement while maintaining clarity and usability. However, identifying and characterizing differences in expressiveness between quantum programming paradigms remains an open area of study. Our work investigates the landscape of expressiveness through a comparative analysis of hosted quantum programming languages such as Qiskit, Cirq, Qrisp, and quAPL, and standalone languages including Q\# and Qmod. We focused on evaluating how different quantum programming languages support the implementation of core quantum algorithms—Deutsch-Jozsa, Simon, Bernstein-Vazirani, and Grover—using expressiveness metrics: Lines of Code (LOC), Cyclomatic Complexity (CC), and Halstead Complexity (HC) metrics as proxies for developer productivity. Our findings suggest that different quantum programming paradigms offer distinct trade-offs between expressiveness and productivity, highlighting the importance of language design in quantum software development.
\end{abstract}


\section{Introduction}
Programming languages are examples of \textit{formal languages}, languages designed to communicate procedural intent between programmers and issue instructions to a machine to perform specific tasks \cite{ari_programming_languages}. These instructions must follow a set of syntactical rules that define how valid ``sentences'' in the language should be structured using a predefined set of primitive symbols and a formal grammar to generate valid expressions. Once the set of syntactically valid instructions is established, semantics assign meaning to these expressions by executing the intended operations when processed by a computing machine. Well beyond their mechanizable purposes, programming languages provide notation that enables thinking. 

In essence, the \textit{syntax} of a programming language dictates the structure of its expressions, statements, and program units, while its semantics define their meaning. Consequently, the choice of programming language imposes certain constraints on software development, as it determines the available control structures, data structures, and abstractions, ultimately shaping the types of algorithms that can be built and how easily programmers can do so \cite{scott_programming_languages_pragmatics}. Following this line of thought, a programming language is more than just a tool for instructing a computer to perform tasks; it also serves as a framework within which we organize our ideas about processes \cite{Abelson1996}.

Currently, classical high-level programming languages abstract hardware specifications, bringing them closer to human cognition. These languages support abstraction mechanisms that ensure a degree of independence from computer organization and architecture, and its physical principles \cite{Gabbrielli2010}. While these languages do continue to evolve, they have reached a significant level of abstraction and maturity, eliminating or reducing the need for programmers to directly engage with low-level implementations. For example, programming at a level below assembly only occurs during microprocessor design phases or firmware updates more regularly.

However, this was not always the case. In its early days, programming a computer was a tedious and constrained task which required working directly with machine language using hardware-specific instructions. The introduction of high-level languages such as FORTRAN in the 1950s marked a turning point \cite{backus1978history} that changed the economics of programming, enabling numerical computations and other operations to be expressed in a more abstract manner closer to human thought. These languages prompted the invention of compilers, tools that functioned as translators between source code and machine code \cite{scott_programming_languages_pragmatics} \cite{chong_programming_quantum_hardware}. Since then, the field of software development has undergone significant evolution, resulting in an extensive array of general-purpose languages, domain-specific languages, diverse programming paradigms, frameworks, tools, and methodologies. These advancements aim to enhance software development productivity, facilitating solutions across most of contemporary human experience from finance and healthcare, to education. In a very precise manner, our digital experience today is a results of having adequate abstraction mechanisms translatable to optimized execution.

Programming \textit{paradigms} describe deliberate choices in code structure and organization, each following specific rules and methods backed by formal rules. Imperative, declarative, functional, and logical paradigms emerged as answers to the question of what form of expressiveness may be more suitable for efficiently solving a problem depending on its nature. Imperative programming focuses on specifying how an algorithm should be implemented. In languages such as FORTRAN, C, C++, Java, or Python, this paradigm enables the construction of algorithms through a sequence of instructions \cite{ari_programming_languages}. Object-oriented programming, with an imperative origin, structures code around objects that encapsulate behaviors and attributes, making it particularly useful for developing complex systems.

In contrast, declarative programming focus on the structure of the problem being solved rather than how to achieve it \cite{scott_programming_languages_pragmatics}. Functional and logic programming belong to this paradigm. Languages such as APL, Lisp, Haskell, and Scheme are  designed around promoting functions as first-class citizens in their type system \cite{sofge_quantum_programming}. In logic programming, computation is driven by facts and logical rules, allowing the system to infer answers incrementally rather than following a predetermined sequence of steps. Instead of specifying how to solve a problem, the system applies logical inference to stated facts and rules to determine solutions. Prolog is a notable example.

Selecting a programming paradigm to solve a problem depends on how well the language form fits the expected functionality within the solution. This choice is not trivial, and it can directly impact code clarity, efficiency, and memory usage. Some problems may be better suited to a specific paradigm, while others may benefit from a combination of multiple paradigms to achieve optimal solutions. With the emergence of quantum computing, the possibility arose that certain problems either difficult or not computable by classical computers could be solved by a quantum computer. This concept was explored by Richard P. Feynman, who proposed that quantum systems could be efficiently simulated only by quantum computers \cite{feymman_simulating_physics}. Despite the evident success of these insights, the preliminary formulation of quantum computing centered on computability and tractability rather than programmability, a feature that arises naturally as quantum hardware becomes possible, and later matures.

Although classical programming languages have reached a high level of maturity, they lack the ability to capture the unique properties of quantum computing such as superposition, entanglement, and interference \cite{b3}. Instead, most quantum programming languages rely on libraries or domain-specific languages (DSLs) to address these challenges and enable quantum programming. Various software development kits (SDKs), frameworks, and quantum programming libraries are hosted in classical programming languages. Qiskit \cite{b7}, PennyLane \cite{b8}, Cirq \cite{cirq_page}, Qrisp \cite{qrisp}, and ProjectQ \cite{projectQ} are examples of tools developed in Python. Quipper, a functional quantum programming language, was implemented in Haskell \cite{quipper}, while quAPL is hosted in APL \cite{quapl}.

Other languages, such as Q\#, take inspiration from classical programming languages --e.g., C\# and F\#--, offering domain-specific constructs for quantum development \cite{qsharp_microsoft, qsharp}. Qmod provides a high-level language with native quantum abstractions, enabling clear algorithmic intent \cite{qmod}. Quil is an instruction-based language designed to express quantum programs with classical control and feedback, serving both as a low-level programming tool and a compilation target for higher-level quantum languages \cite{quil}. Scaffold, on the other hand, adopts an imperative programming model with C-like syntax, combining classical control structures and modular design to express quantum functionality. \cite{scaffold}.

From the perspective of quantum algorithm research, our understanding of quantum computing is grounded in the mathematical principles that underpin it as a first-order approximation. These principles define how we reason about quantum algorithms and their implementation, which at their most fundamental level are equivalent to linear algebra on Hilbert spaces with complex support, with linear logic providing the context of how to reason with non-reusable resources \cite{valiron2008semantics}. However, these complexities challenge the expressiveness of quantum programming languages, which rely on classical foundations but lack the native abstractions necessary to fully harness quantum mechanics; quantum programming is a hard task today \cite{di2024abstraction}. This limitation impacts how we conceptualize and implement quantum algorithms, as well as the likelihood of producing correct programs from the outset.

Given the above, our work aims to uncover properties of quantum problems and quantum programming languages that maximize the likelihood of reliably producing correct programs. By exploring the expressiveness of quantum languages and their alignment with quantum algorithms, our goal is to bridge the gap between theoretical principles and practical implementation. To achieve this, our research evaluates the expressiveness of quantum programming languages through a comparative analysis of quantum algorithms in hosted and standalone languages, drawing  comparisons between both.

We report here the results of an evaluation comparing the expressiveness and productivity of hosted quantum programming languages like Qiskit, Cirq, Qrisp, and quAPL, and standalone languages, such as Q\# and Qmod. The focus of this work is on how these languages enable the abstraction of physical concepts for quantum computing through the implementation of the Deutsch-Jozsa, Simon, Bernstein-Vazirani, and Grover algorithms. We conducted our assessment using a set of metrics, including Cyclomatic Complexity (CC), Lines of Code (LOC), and Halstead Complexity metrics (HC)\cite{metrics_complexity}, all of them grounded in classical programming language theory. Together, these metrics evaluate syntactic expressiveness, semantic richness, and lexical structure by analyzing control flow, code length, and the frequency and variety of operators and operands used in the implementation of quantum algorithms. As part of this comparative study, our research aims to evaluate the expressiveness of quantum programming languages and their impact on the construction of productive quantum programming units.

\section{Metrics for Expressiveness Evaluation}

Expressiveness is a key property of programming languages, directly linked to the grammar that defines them, the paradigms they support, and the underlying abstract machine on which they compute \cite{b5}. As a result, the choice of a programming language influences the complexity of algorithm implementation, as it directly affects how problems are conceptualized \cite{heim_quantum}. The \textit{expressive power} of a language measures the range of ideas that can be described within it and is strongly influenced by the constructs it provides. It is also related to the ability of a programming language to represent complex ideas and algorithms in a clear and concise manner. This means certain problems may require significantly more implementation effort in some languages compared to others \cite{b12}.

Among various dimensions of expressiveness, \textit{syntactic expressiveness} plays a central role, as it relates to the ability to represent concepts structurally while doing so economically. This has been studied through the analysis of reductions between languages that preserve operational equivalence, providing a means of assessing how abstract and compact a language can be \cite{b12, b18}. A key aspect of this analysis is \textit{observational equivalence}, which holds that two programs or expressions are considered equivalent if, in all possible contexts, they produce the same observable results \cite{b13, b18}. Furthermore, syntactic expressiveness entails transforming constructs from one language to another while preserving their observational semantics \cite{b18}.

Expressiveness relates to code conciseness. The \textit{Conciseness Conjecture} introduced by Felleisen states that more expressive languages tend to yield more compact programs by offering powerful semantic constructs that reduce the need for repetitive or explicit structures \cite{b29}. Davidson and Michaelson examine this conjecture empirically, showing that computational models with richer semantics produce shorter representations for functionally equivalent algorithms \cite{b14}. This theoretical and empirical foundation supports the use of structural metrics to study expressiveness quantitatively, which constitutes the basis of our study and results below.

To provide a rigorous and quantifiable evaluation of expressiveness, our study focuses on structural complexity metrics intended to measure the intellectual efficiency of producing and maintaining quantum programs.  Our study applied three classical software complexity metrics: Lines of Code, Cyclomatic Complexity, and Halstead Complexity. Each captures different aspects of program structure, and when combined, they provide a more complete view of code clarity, abstraction, and implementation effort. While no single metric can fully describe software complexity --or the complexity of its production-, existing literature supports using a combination of complementary measures to gather better supporting evidence during comparative analyses \cite{metrics_complexity}. These metrics are independent of a language and allow for static analysis of code. When applied in combination with other metrics such as Cyclomatic Complexity and LOC, Halstead metrics help quantify expressiveness in terms of symbolic structure and implementation effort.

\subsection{Lines of Code (LOC)}
LOC measures the number of source code lines in a program. It serves as a baseline indicator of code verbosity and can reflect the syntactic expressiveness of a language and helps to compare how different quantum programming languages express equivalent algorithms \cite{b12}. Although LOC does not capture control flow, logical complexity, or code semantics, it remains useful for identifying how concisely a language expresses an algorithm. \cite{metrics_complexity}.

\subsection{Cyclomatic Complexity (CC)} A widely used software metric introduced by Thomas McCabe, which measures the number of linearly independent paths in a program. A higher CC value indicates increased control flow complexity, which can make debugging and verification more challenging \cite{b19, metrics_complexity}. In quantum programming, CC also signals how much classical control logic is needed to express quantum algorithms \cite{yuan2024quantum}, which can impact readability and verification.

\subsection{Halstead Complexity (HC)}
Halstead’s software metrics evaluate programs based on their lexical structure. It measures program volume, difficulty, and effort using the number of distinct and total operators and operands. These values seek to approximate the cognitive load required to understand, implement, or modify code. Despite Halstead metrics not accounting for control structures or semantic correctness, they provide valuable insights into the symbolic density and structural complexity of a program\cite{metrics_complexity}.

\begin{itemize}
    \item $n_1$: Number of distinct operators
    \item $n_2$: Number of distinct operands
    \item $N_1$: Total number of operators
    \item $N_2$: Total number of operands
\end{itemize}

From these values, Halstead defined several derived metrics:

\paragraph{Program Vocabulary} 
Total number of unique symbols (operators and operands):
\begin{equation} n = n_1 + n_2  \end{equation}

\paragraph{Program Length} 
Total number of lexical tokens:
\begin{equation}   N = N_1 + N_2 \end{equation}

\paragraph{Volume (V)} 
Size of the implementation in terms of information content:
\begin{equation}   V = N \cdot \log_2 n \end{equation}
A higher volume indicates more information is encoded in the program, which may imply greater effort to comprehend or maintain it.

\paragraph{Difficulty (D)} 
Cognitive difficulty of understanding or writing the program:
\begin{equation} D = \frac{n_1}{2} \cdot \frac{N_2}{n_2} \end{equation}
This metric increases with more complex or less reusable code elements.

\paragraph{Effort (E)} 
Total mental effort required to develop or understand the code:
\begin{equation}  E = D \cdot V \end{equation}
Effort correlates with development time and is often used as an indicator of programming complexity.

\section{Landscape of Quantum Programming Languages}

We describe here quantum programming languages selected for analysis. Our selection comprises both hosted and standalone languages that reflect a diverse range of paradigms, abstraction mechanisms, and design mechanisms. These languages --Qiskit, Cirq, Qrisp, and quAPL (hosted), along with Q\# and Qmod (standalone)-- were chosen for their active development, open-source availability, and distinctive methods to represent quantum algorithms. These languages differ in how they expose circuit construction, state manipulation, and quantum-classical interaction. By analyzing them in a unified framework, we aim to surface structural characteristics that affect productivity, clarity, and algorithmic scalability. We seek to understand how each language models quantum computation, which abstractions they provide, and how their underlying design influences code expressiveness. 

\subsection{Hosted Languages}
\paragraph{Qiskit}
Qiskit is a quantum software development framework developed by IBM that enables users to create and manipulate quantum circuits, simulate their execution, error correction, and integration on actual quantum hardware. It is implemented as a Python-based hosted language, leveraging the accessibility and ecosystem of Python to facilitate quantum programming \cite{qiskit}. Qiskit follows the imperative paradigm, where quantum algorithms are explicitly constructed by defining quantum gates and operations in a sequential manner. The design of Qiskit emphasizes a low-level, gate-based design, giving users direct control over quantum circuit construction. This level of control makes it suitable for precise algorithm implementation, although it may lead to increased verbosity and complexity in code due to the need for detailed specification of operations.

\paragraph{Cirq}
Cirq is a quantum programming framework developed by Google, designed for the construction, simulation, and execution of quantum circuits, targeting near-term quantum computers \cite{cirq_page}. It is a Python-based hosted framework that emphasizes fine-grained control over quantum operations, providing tools to define, manipulate, and optimize quantum circuits at the level of individual quantum gates. Cirq is also imperative: users explicitly construct circuits by sequencing gate operations over qubits. Circuits in Cirq can be represented using a \texttt{Circuit} object or a \texttt{Schedule} object, the latter offering greater control over the timing and alignment of operations. This abstraction allows developers to design circuits with precise execution order and synchronization, allowing optimized circuit execution on real quantum devices \cite{cirq_page_2}.

\paragraph{Qrisp}
Qrisp is an embedded domain-specific language (eDSL) written in Python. The structured programming paradigm of Qrisp simplifies the development and maintenance of scalable quantum programs. The language integrates seamlessly into classical Python code, allowing developers to define quantum types that can be used like any standard Python variable \cite{qrisp}. The core abstraction in Qrisp is the \texttt{QuantumVariable}, which hides qubit management from the user and enables human-readable inputs and outputs. Qrisp allows the definition of arithmetic expressions and control flow statements that operate directly on quantum variables. It supports conditionals and loops applied to quantum registers, and manages low-level circuit transformations such as uncomputation and ancillary qubit cleanup internally. By generating fully compilable circuits from high-level symbolic code, Qrisp aims to bridge the gap between quantum algorithm design and practical hardware implementation \cite{qrisp}.

\paragraph{quAPL}
quAPL is a quantum programming language embedded in APL that models quantum computation using linear algebra and array-based constructs. It removes the need for explicit gate-level definitions by allowing the definition of \emph{quantum motifs} that encapsulate patterns of computation, executable then efficiently at the level of linear algebra \cite{quapl}. The language uses algebraic composition to express quantum behavior with specialized types for quantum states and operations such as unitary application, tensor products, and measurement. This way of expressiveness enables compact, symbolic, and high-level representations of quantum logic. quAPL inherits its declarative and functional form from APL, emphasizing mathematical transformations and combination of functions and operators over control flow. Rather than constructing explicit circuits, quAPL composes transformations over vector states. 

\subsection{Standalone Languages}

\paragraph{Q\#}
Q\# is a standalone domain-specific language developed by Microsoft for expressing quantum algorithms, with a hybrid functional-imperative syntax offering structured abstractions specifically designed for quantum computation \cite{qsharp, qsharp_microsoft}. Q\# encourages algorithmic thinking over circuit construction by allowing quantum operations to be combined with classical control flow structures such as conditionals, loops, and repeat-until-success constructs. Q\# separates classical logic from quantum operations by having classical behavior implemented via \texttt{function} declarations, while quantum procedures are expressed as \texttt{operations} that act on \texttt{qubit} types. It also introduces intrinsic quantum constructs such as \texttt{adjoint} and \texttt{controlled} functors, and provides additional features like \texttt{mutable} variables, algebraic types, and type inference. Qubit management is handled explicitly through \texttt{using} and \texttt{borrowing}, which help track lifetime and reuse. This design supports composability, resource safety, and correctness in quantum algorithm development.

\subsubsection{Qmod}
Qmod is a standalone high-level quantum programming language designed to express quantum programs through composable, declarative constructs. Its design emphasizes algorithmic intent rather than low-level control, allowing developers to describe quantum behavior using human-readable abstractions \cite{qmod}. The language introduces a set of core language constructs such as \texttt{repeat}, \texttt{within}, \texttt{apply}, \texttt{lambda}, and \texttt{let}, which support iteration, block structure, functional composition, and scoped variable definitions. Quantum operations are represented through intuitive primitives like \texttt{init}, \texttt{unitary}, \texttt{entangle}, and \texttt{measure}, allowing the programmer to focus on the algorithm’s semantics rather than circuit wiring. Qmod is functional and declarative, and its syntax separates quantum logic from classical control flow. It also supports modular composition and type inference, promoting expressiveness and reusability. By abstracting away the underlying circuit structure, Qmod seeks to enable  developers by focusing on scalable quantum programs while preserving clarity and correctness.

\section{Overview of the Experimental Setup}  

To evaluate the expressiveness of quantum programming languages, we conducted a series of experiments using well-established expressiveness metrics described above (i.e., Lines of Code, Cyclomatic Complexity, Halstead Complexity). In this study, we hypothesize that, to a first approximation, these selected metrics provide quantitative and relevant evidence of how different languages enable the representation of quantum algorithms similar to how they operate in classical code. Our study compares \textit{hosted} and \textit{standalone} quantum programming languages under the same experimental framework to ensure consistency in evaluation. Hosted languages (i.e., Qiskit, Cirq, Qrisp, quAPL) rely on classical programming languages (i.e., Python, APL) for implementation, whereas standalone languages (i.e., Q\#, Qmod) are designed with quantum-native constructs.  

\subsection{Selection of Quantum Algorithms} 

To perform these experiments, we selected four well-known quantum algorithms. Our choice is given by a) how well known these are in the quantum programming community, b) common execution in quantum hardware, c) frequency of appearance in the introductory steps of learning about quantum programming, d) degree of presence of superposition, interference and entanglement, and e) direct connection to modes of reasoning behind how quantum algorithms operate generally. Each algorithm was implemented in each quantum programming language and evaluated with the same metrics. Our results are available on GitHub for reproducibility purposes
\footnote{See:\url{https://github.com/Universidad-Cenfotec/quantum-programming-languages-expressiveness}.}.

\begin{itemize}
    \item \textbf{Deutsch-Jozsa:} Each implementation includes the full circuit design, oracle functions, and the algorithm for both constant and balanced functions. The evaluation used 3 qubits plus 1 ancillary qubit.  
    \item \textbf{Bernstein-Vazirani:} Evaluated with 4 qubits plus 1 ancillary qubit and a fixed hidden bitstring `1101`.  
    \item \textbf{Simon's Algorithm:} Implemented using 3 input qubits plus 3 ancillary qubits, with a hidden bitstring `101`.  
    \item \textbf{Grover's Search:} Evaluated with 3 qubits, performing two iterations, where the marked state was `101`.  
\end{itemize} 

\subsection{Implementation and Code Analysis}  
We developed a Python project to analyze the source code files and remove non-essential lines, including blank lines and commented lines. To measure Cyclomatic Complexity, we used McCabe's metric, but applied it at the whole program level rather than to individual functions. Unlike traditional, per-function CC analysis, our implementation evaluates the entire circuit design, which includes the complete quantum algorithm implementation and any oracles used by the algorithm. The CC calculation added one to the total to reflect the number of independent paths within the script. Table~\ref{tab:cc_constructs} summarizes the constructs considered in each language to account for control flow complexity within CC.

\begin{table}[h]
\centering
\caption{Programming Constructs Used in CC Analysis}
\label{tab:cc_constructs}
\renewcommand{\arraystretch}{1.5}
\begin{tabular}{c p{5cm}}  
\hline
\textbf{Language} & \textbf{Constructs Considered for CC Analysis} \\ \hline
Qiskit, Qrisp, Cirq (Python) & if, elif, for, while, except, with, assert, list, set, dict, and, or, on\_each \\ 
quAPL (APL) & \includegraphics[height=20pt]{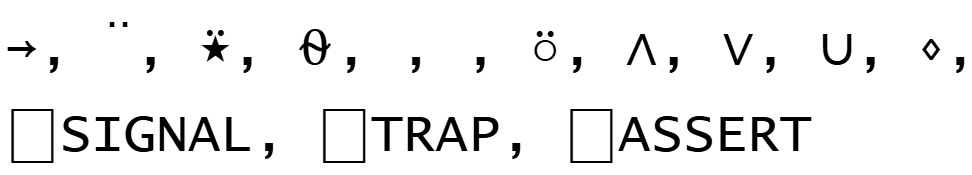} \\ 
Qmod & if, else, repeat, within, apply, lambda, and, or, on\_each \\ 
Q\# & if, elif, for, ApplyToEachA,MeasureEachZ, try, catch, repeat, until \\ 
\hline
\end{tabular}
\end{table}

\subsection{Scope and Limitations}

Although our data capture and analysis tool was designed to be scalable, we focus here only on LOC, CC and Halstead metrics. We anticipate future quantum-native code evaluation metrics may arise, thus the extensible design of our tooling. This study is limited to the languages mentioned in the analyzed literature, so other quantum programming languages that were not included remained outside the scope of this research, although they could be considered in future studies.

We chose four fundamental algorithms in order to consistently build the evaluation methodology with compact yet meaningful examples. The proposed methodology may be similarly expanded to include more algorithms in future exercises. We are aware that our choice of algorithms only consists of the most simple ones, which may bias the results to a subset of possible algorithmic patterns.

Most of the implementations were coded from scratch, as was the case with Qiskit, Cirq, Qrisp, and Q\#. Meanwhile, others were taken directly from libraries provided by developers (Qmod and quAPL). We are aware that the quality of the implementations we used for the assessment correspond to those produced by intermediate users, and this fact may also bias our results. At the same time, and given the state of the quantum community, this is not an unreasonable assumption to make.

The efficiency of algorithm execution was not considered, as the study focused on the syntactic and structural representation of languages rather than their performance in terms of execution time or resource consumption. Consequently, we did not evaluate performance on real quantum hardware, as the focus was on code expressiveness and its implementation in the design of quantum algorithms rather than execution. 

\section{Results}
To quantify the expressiveness of quantum programming languages, we applied our tools to measure Lines of Code, Cyclomatic Complexity, and Halstead Complexity across multiple implementations of four quantum algorithms. The following figures and tables report the key findings of our study.

\subsection{Lines of Code Analysis}
\begin{figure*}[h]
    \centering
    \includegraphics[width=0.8\linewidth]{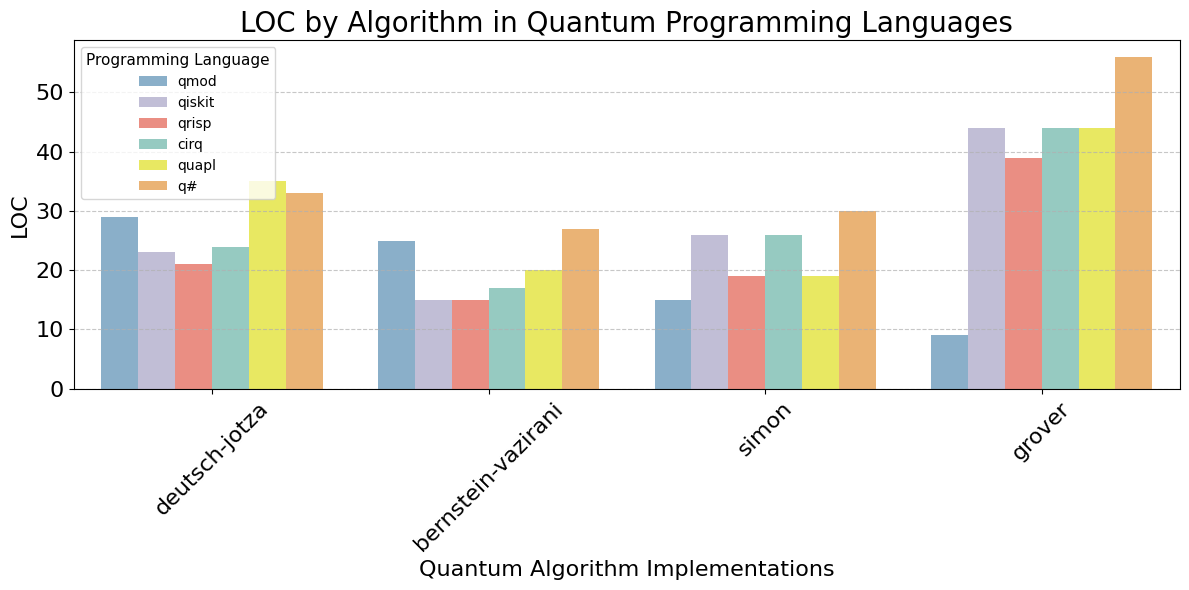}
    \caption{Lines of Code across quantum algorithms and quantum programming languages.}
    \label{fig:loc_all_qpl}
\end{figure*}

The comparison of LOC across different quantum programming languages provides insight into syntactic expressiveness and their verbosity (Table~\ref{tab:loc}). These results provide grounds for a direct comparison of the LOC required for each algorithm across different quantum programming languages. We deaggregated LOC data, per algorithm implementation across all languages to visualize variation (Fig.~\ref{fig:loc_all_qpl}). Grover's algorithm typically requires the most lines of code. Most languages implement Grover with the highest LOC, which is expected due to its complexity compared to the other evaluated algorithms. However, Qmod behaves unusually, as it requires the fewest lines (9), even fewer than Deutsch-Jozsa (29) or Simon (15). This pattern suggests that Qmod includes specific abstractions for Grover, reducing the amount of code needed.

\begin{table}[h]
    \centering
    \caption{LOC Scores for Quantum Algorithms}
    \label{tab:loc}
    \renewcommand{\arraystretch}{1.2}
    \begin{tabular}{lcccc} 
        \hline
        \textbf{Language} & \textbf{Deutsch-Jozsa} & \textbf{Bernstein-Vazirani} & \textbf{Simon} & \textbf{Grover} \\ \hline
        Cirq     & 24 & 17 & 26 & 44 \\
        quAPL     & 35 & 20 & 19 & 44 \\
        Qiskit   & 23 & 15 & 26 & 44 \\
        Qrisp    & 21 & 15 & 19 & 39 \\
        Qmod     & 29 & 25 & 15 & 9  \\
        Q\#      & 33 & 27 & 30 & 56 \\
        \hline
    \end{tabular}
\end{table}
\vspace{0.8em}
\noindent

Cirq, Qiskit, and Qrisp share similar results. These three languages have closely matching LOC across all implementations. Since they share Python as their host language, they likely follow similar code structures and rely on comparable libraries that optimize algorithm implementations uniformly. 

Q\# consistently requires more lines of code than any other language. For example, it reaches 33 LOC for Deutsch-Jozsa and 56 for Grover. This suggests that its modular architecture and strong static typing introduce additional verbosity in algorithm construction, especially in oracle definitions and measurement logic.

quAPL, also demonstrates consistently high LOC across algorithms. Despite the syntactic brevity of APL, quAPL records 35 LOC for Deutsch-Jozsa, 20 for Bernstein-Vazirani, and 44 for Grover. This indicates that while quAPL may use dense expressions, its functional style still demands structural clarity that translates into longer code.

Qmod stands out for its highly compact implementation of Grover’s algorithm (9 LOC), which is noticeably lower than its implementations of Bernstein-Vazirani (25 LOC) and Deutsch-Jozsa (29 LOC). This difference suggests that Qmod includes specific abstractions optimized for Grover, whereas other algorithms may still require more explicit or verbose logic.

\begin{figure}[h]
    \centering
    \includegraphics[width=\linewidth]{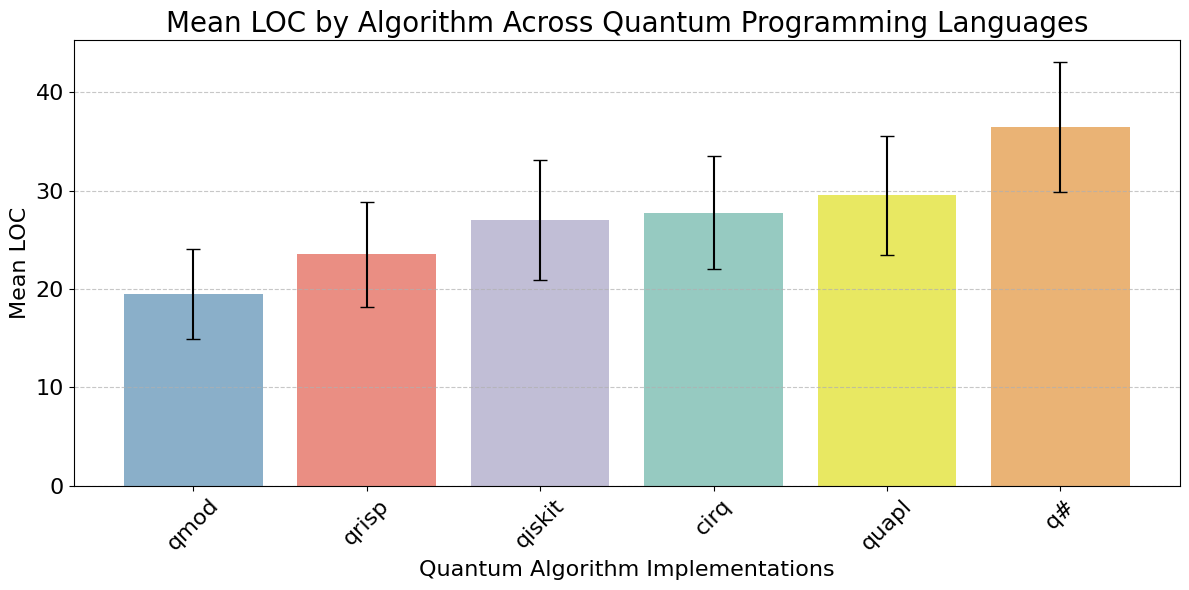}
    \caption{Mean LOC across quantum programming languages.}
    \label{fig:loc_avg}
\end{figure}

These results indicate that both languages define different routes to code expressiveness and structure. Despite being a crude metric, average LOC per language are suggestive of differences in overall verbosity across languages independent of specific algorithms (Fig.~\ref{fig:loc_avg}). Qmod exhibits the lowest average number of lines of code. This language maintains the most compact implementation of the evaluated algorithms, suggesting the presence of efficient abstractions to reduce code verbosity. By contrast, Qrisp, Qiskit, Cirq, and quAPL have similar average LOC values. These three languages fall into an intermediate position in terms of verbosity, sharing a similar level of expressiveness when implementing quantum algorithms.

\subsection{Cyclomatic Complexity Analysis}

\begin{figure*}[h]
    \centering
    \includegraphics[width=0.8\linewidth]{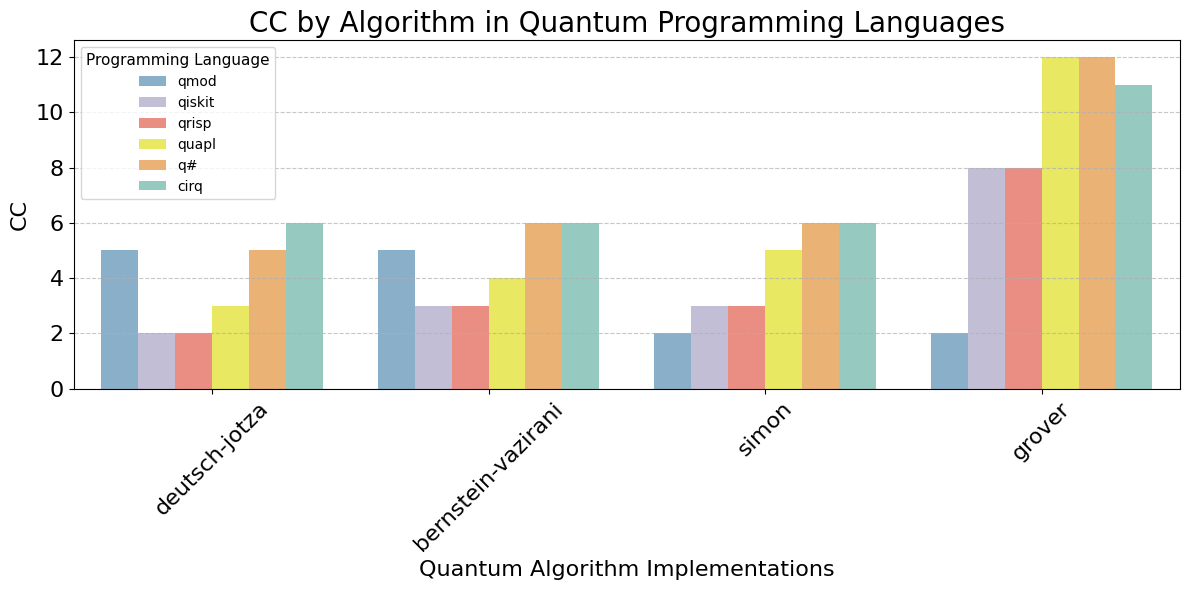}
    \caption{CC across quantum algorithms and quantum programming languages.}
    \label{fig:cc_all_qpl}
\end{figure*}

We evaluated cyclomatic complexity scores for each algorithm and quantum programming language (Table~\ref{tab:cyclomatic_complexity}). Obtained data provide various insights on the effect of language features into the structural complexity of resulting implementations. We also visualized CC scores per algorithm implementation across languages (Fig.~\ref{fig:cc_all_qpl}) to better highlight structural differences.

\begin{table}[h]
    \centering
    \caption{Cyclomatic Complexity Scores per Quantum Algorithms and Quantum Programming Language}
    \label{tab:cyclomatic_complexity}
    \renewcommand{\arraystretch}{1.2}
    \begin{tabular}{lcccc} 
        \hline
        \textbf{Language} & \textbf{Deutsch-Jozsa} & \textbf{Bernstein-Vazirani} & \textbf{Simon} & \textbf{Grover} \\ \hline
        Cirq     & 6 & 6  & 6 & 11 \\
        quAPL     & 3  & 4 & 5  & 12 \\
        Qiskit   & 2 & 3  & 3 & 8 \\
        Qrisp    & 2 & 3  & 3 & 8 \\
        Qmod     & 5 & 5  & 2  & 2  \\
        Q\#      & 5  & 6  & 6  & 12 \\
        \hline
    \end{tabular}
\end{table}

Grover remains the algorithm with the highest cyclomatic complexity in most of the evaluated languages. Q\# and quAPL with a score of 12, and Cirq with 11, have the most complex implementations in terms of CC. These values suggest that these languages require a significant number of control structures to implement Grover, which aligns with their behavior in terms of lines of code. However, Qmod behaves unusually, with a score of only 2 for Grover and Simon, the lowest among all languages. This result is particularly notable because its cyclomatic complexity for Deutsch-Jozsa and Bernstein-Vazirani is higher, with values of 5. This suggests Qmod includes abstractions in its library that privilege efficient representation of transformations appearing in the algorithm.

Qrisp and Qiskit follow similar trends across all algorithms, with complexity scores that remain closely aligned. This finding reinforces the idea that both languages, by sharing Python as their host language, may adopt similar strategies for structuring code and implementing quantum algorithms. The complexity in their scores for Deutsch-Jozsa (2 for both), Bernstein-Vazirani (3 each), Simon (3 each), and Grover (8 for Qiskit, 8 for Qrisp) reflect this consistency.
quAPL scores range from 3 to 12, with particularly intermediate values in Bernstein-Vazirani and Simon (5).
Notably, quAPL presents some of the highest individual cyclomatic complexity scores, particularly in  Grover (12). While its overall complexity is not consistently the highest across all algorithms, these peaks suggest that APL’s terse and expressive style may lead to dense implementations that involve non-trivial control flows in certain cases. Cirq and Q\# exhibit the highest average cyclomatic complexity among all evaluated languages, both with mean scores of 7.25 across the four algorithms. This suggests that, while both languages offer rich control-flow mechanisms and abstraction features, their implementations may involve more branching and conditional logic. 

\begin{figure}[h]
    \centering
    \includegraphics[width=\linewidth]{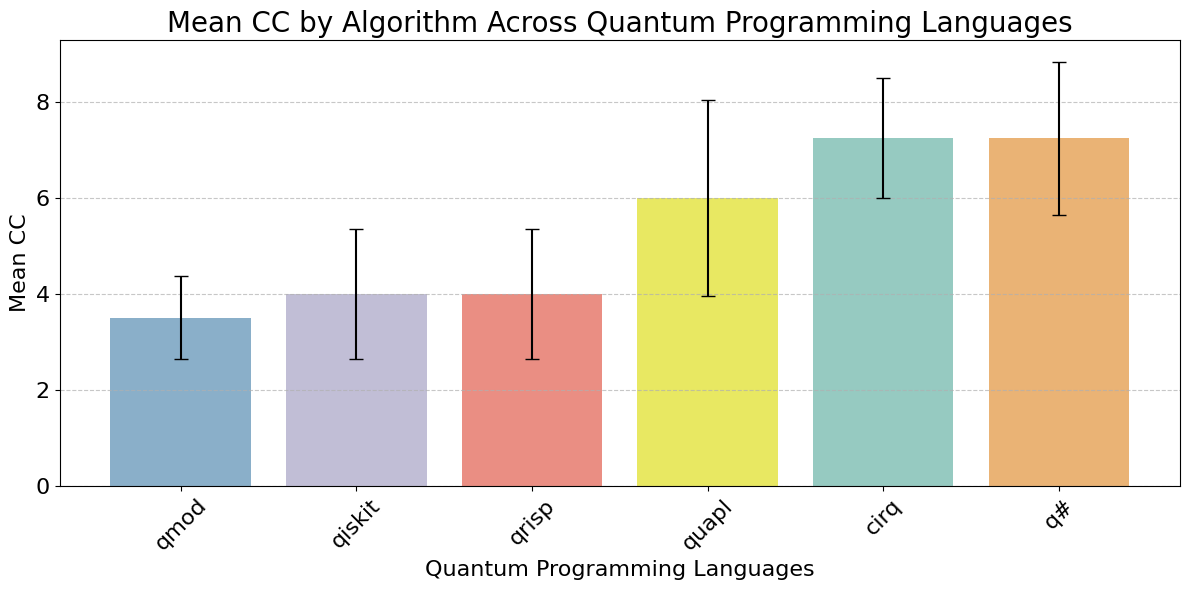}
    \caption{Mean CC across quantum programming languages.}
    \label{fig:cc_avg}
\end{figure}

Overall, cyclomatic complexity varies significantly across languages, showing patterns consistent with previous findings on lines of code. Grover remains the most complex algorithm in most cases, with the exception of Qmod, which notably breaks this trend. Cirq and Q\# exhibit the highest average complexity, reflecting more extensive use of control structures and branching. In contrast, Qiskit and Qrisp maintain low and nearly identical complexity scores, suggesting streamlined implementations possibly influenced by Python’s expressive syntax. quAPL shows moderate to high complexity in some algorithms, particularly Grover.
These results reinforce the idea that the choice of quantum programming language affects not only the size of the codebase but also the structural intricacy of the implementation. These trends are further supported by Figure~\ref{fig:cc_avg}, which illustrates the average cyclomatic complexity per language. The highest complexity levels are observed in Cirq and Q\#, while Qmod stands out for offering the most compact control flow across all evaluated cases.

\begin{figure*}[t]
    \centering
    \includegraphics[width=\textwidth]{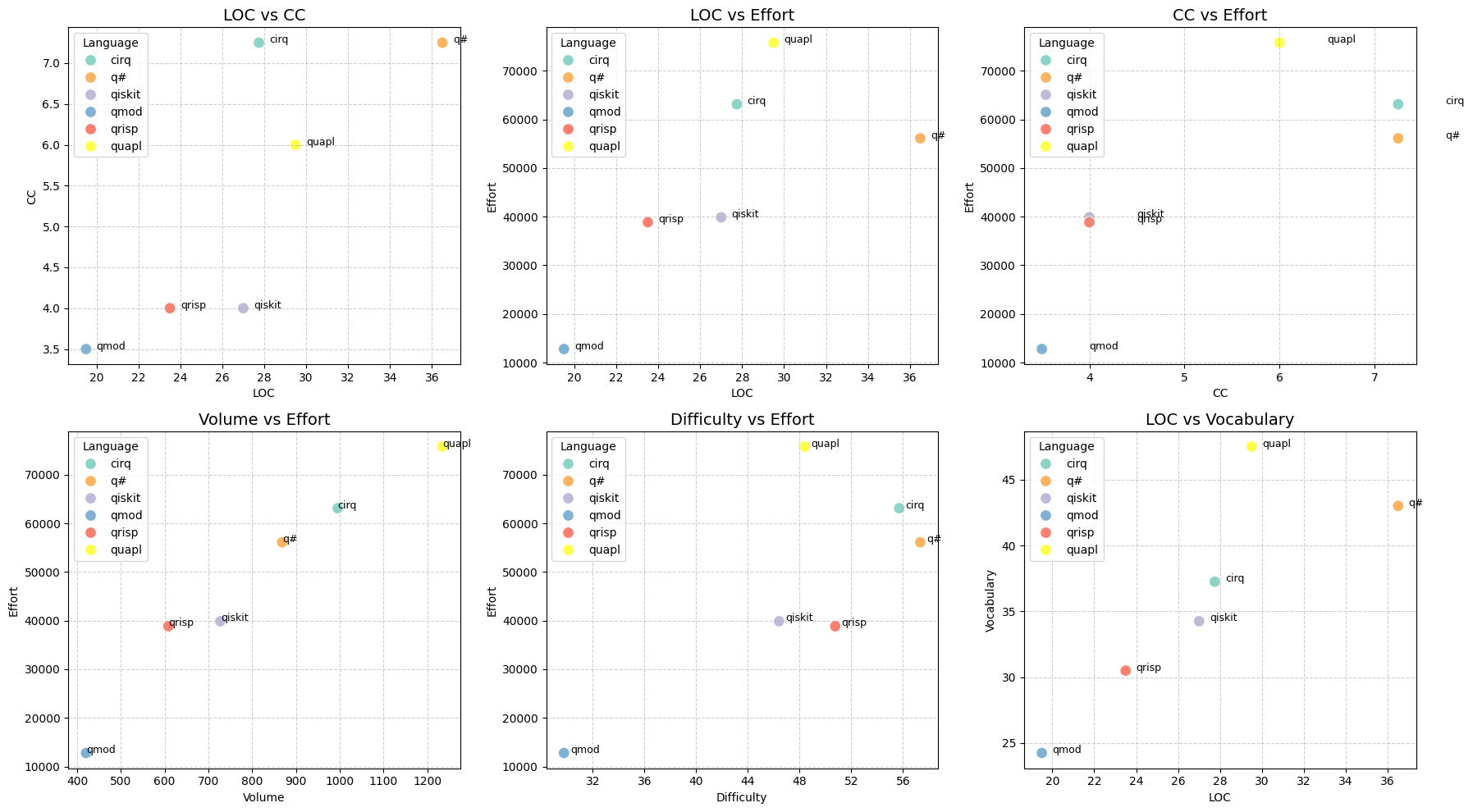}
    \caption{Relationship between complexity metrics applied to quantum programming languages. Each point corresponds to the mean per language for all algorithms.}
    \label{fig:scatter_metrics}
\end{figure*}

\subsection{Complexity Metrics Analysis}

\begin{figure}[h]
    \centering
    \includegraphics[width=0.8\linewidth]{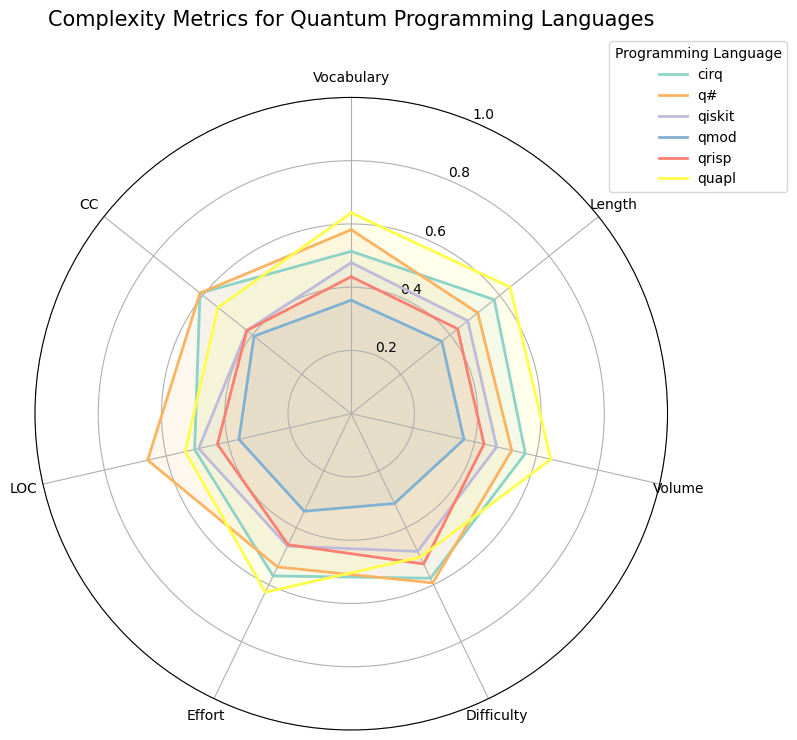}
    \caption{Metrics comparing quantum programming languages.}
    \label{fig:complexity_metrics_radar}
\end{figure}

Given the richness of the Halstead Complexity metrics, we decided to evaluate the relationship between Lines of Code (LOC), Cyclomatic Complexity (CC), Effort, Volume, Difficulty, and Vocabulary, following the definitions by Halstead and McCabe (Fig. ~\ref{fig:scatter_metrics}). Additionally, Fig. ~\ref{fig:complexity_metrics_radar}  summarizes the findings reported below.

\paragraph{LOC vs CC}
We found some degree of separation between languages that favor high-level abstractions and those that rely on more explicit control structures. Q\# and Circ stands out with the highest lines of code and cyclomatic complexity, reflecting a programming model that emphasizes explicit quantum operation declarations and strict type enforcement. quAPL also exhibits high LOC and CC, nearly on par with Cirq, despite being based on APL, a language known for terse syntax.

In contrast, Qiskit and Qrisp cluster together in the lower-left region of the plot, displaying both low LOC and low CC. This pattern can be attributed to the use of array-based abstractions and internal iteration mechanisms. For instance, in Qrisp, operations like \texttt{qrisp.h(input)} or \texttt{qrisp.cx(a, b)} are implicitly broadcasted over quantum registers, eliminating the need for explicit \texttt{for} loops in most quantum algorithms. Similarly, Qiskit often utilizes Pythonic range-based constructs and functional programming patterns that reduce syntactic overhead. These abstractions help streamline circuit construction and reduce branching, which directly impacts cyclomatic complexity. Finally, Qmod consistently appears as the most compact language in both metrics, indicating a declarative and minimalistic approach that simplifies both the structural and syntactic layers of quantum program design.

\paragraph{LOC vs Effort}
quAPL  shares a similar place with Cirq in having the  highest effort values. Q\# seems to follow a somewhat linear trend between LOC and effort alongise Qiskit, Qmod and Qrisp . Qmod remains at the lower end for both LOC and effort, suggesting minimal cognitive load in proportion to code volume.

\paragraph{CC vs Effort}
quAPL, Cirq, and Q\# occupy the upper end of both cyclomatic complexity and Halstead effort, suggesting high structural and cognitive demands. quAPL reaches the highest effort overall, likely due to the dense lexical nature of APL, which, despite its brevity, encodes significant semantic complexity. Cirq follows with similarly high scores, reflecting its low-level, imperative design. At the opposite end, Qmod shows the lowest complexity and effort, reinforcing its compact style. Qiskit and Qrisp form a middle ground: both maintain low CC, but moderate effort, likely due to their use of expressive abstractions over Python’s syntax, which reduce control structures but still entail cognitive overhead.

\paragraph{Volume vs Effort}
All evaluated languages fit a linear trend between volume (information content) vs effort. Qmod remains as the most compact and easily interpretable in our evaluation set. Despite being symbolic compact quAPL appears to force program writers to unpack large information contents at high mental effort; surprisingly, Cirq does as well despite belonging to a different programming paradigm. The remaining languages appear to show metrics determined by the absence (Qiskit, Qrisp) or presence (Q\#) of static type checking.

\paragraph{Difficulty vs Effort}
Qmod bears a striking advantage, favoring both writing and reasoning about modular code, as well as reading the resulting program. Other languages sit at considerable distance from it (higher difficulty). Once again, static typing may be responsible for the position Q\# occupies in the space. Qiskit and Qrisp once again appear to be similar. quAPL and Cirq appear to be both expensive to write programs with and expensive to read quantum programs in.

\paragraph{LOC vs Vocabulary}
APL is known for its lexical richness, which shows in its large vocabulary, sitting at a midpoint from other languages with Circ in terms of LOC. The remaining languages follow a somewhat linear trend, with Q\# at the right-most extreme in LOC for similar reasons as above. Qmod at the lower left.

\begin{table*}[h]
    \centering
    \caption{Average Halstead Metrics by Quantum Programming Language}
    \label{tab:halstead_updated}
    \renewcommand{\arraystretch}{1.2}
    \begin{tabular}{lcccccc} 
        \hline
        \textbf{Metric} & \textbf{Cirq} & \textbf{Q\#} & \textbf{quAPL} & \textbf{Qiskit} & \textbf{Qmod} & \textbf{Qrisp} \\ \hline
        n\textsubscript{1} (Unique Operators)   & 24.00  & 29.25 & 28.50 & 20.75 & 15.50 & 20.00 \\
        n\textsubscript{2} (Unique Operands)    & 13.25  & 13.75 & 19.00 & 13.50 & 8.75  & 10.50 \\
        N\textsubscript{1} (Total Operators)    & 127.50 & 105.75 & 152.50 & 79.75 & 59.25 & 71.25 \\
        N\textsubscript{2} (Total Operands)     & 61.50  & 52.75 & 65.75 & 60.25 & 32.75 & 50.00 \\
        Vocabulary                              & 37.25  & 43.00 & 47.50 & 34.25 & 24.25 & 30.50 \\
        Length                                  & 189.00 & 158.50 & 218.55 & 140.00 & 92.00 & 121.25 \\
        Volume                                  & 995.22 & 868.27 & 1234.53 & 727.26 & 420.47 & 608.55 \\
        Difficulty                              & 55.71  & 57.34 & 48.40 & 46.41 & 29.76 & 50.76 \\
        Effort                                  & 63116.99 & 56107.02 & 75780.32 & 39869.05 & 12781.16 & 38853.50 \\
        \hline
    \end{tabular}
\end{table*}

\section{Discussion}
Results from comparative analysis of quantum programming languages reveals structural and lexical differences that directly impact implementation complexity. In short, these metrics tell a story of expected and unexpected regularities across the expressive power of six quantum programming languages. In particular, our research shows that the divide between hosted and standalone languages is not necessarily clear cut, and that more factors seem to be at play.

Overall, the results show varying relationships between LOC and CC across languages. While Qiskit and Qrisp maintain both metrics at low levels, likely due to their use of high-level abstractions, Cirq exhibits high cyclomatic complexity despite moderate LOC, indicating a more granular and imperative coding style, even though it is also based on Python. Q\# sits at the upper end of both dimensions, reflecting verbose and structurally complex implementations. Halstead metrics further distinguish languages by capturing lexical density and cognitive effort. Q\#

Qrisp and Qiskit show strong alignment in both LOC and CC metrics. These languages share a common host environment (Python), which may explain their structural similarity and predictable behavior. Their results display similar patterns across the board, reinforcing the idea that they follow a shared modular and syntactic design. Qrisp consistently occupies a middle ground across our experiments. Its LOC and CC values are lower than those of Cirq or Qiskit but higher than Qmod. This balance between conciseness and structural control suggests a design philosophy that avoids extremes in verbosity or abstraction. Qmod records the lowest LOC and CC values, along with minimal effort and difficulty scores based on Halstead's definitions. 

Qmod stands out for its concise and declarative syntax, which enables the expression of complex logic using compact functional constructs. As illustrated by its use of inline lambda functions and nested logical operators, for example, in Grover’s algorithm, Qmod avoids explicit control structures in favor of composable quantum primitives. This design contributes directly to its minimal cyclomatic complexity and low Halstead effort across all evaluated algorithms. Its support for functional programming appears to streamline circuit construction while reducing structural and syntactic overhead, making it a compelling language for efficient quantum programming.

On the other hand, quAPL exhibits an atypical profile. Despite its syntactically concise nature, it shows high cyclomatic complexity, effort, and difficulty scores. This suggests that quAPL implementations, while brief in appearance, demand substantial cognitive processing due to symbolic density and lexical variety. Notably, both Qmod and quAPL support functional programming constructs, which may contribute to reduced reliance on explicit branching. However, only Qmod consistently translates these constructs into lower structural and lexical complexity, while quAPL retains a high semantic load.

\section{Conclusions}
This study examined the expressiveness and productivity of quantum programming languages through structural complexity metrics applied to representative quantum algorithms. Our analysis provided quantitative evidence of how language design—paradigms, abstraction mechanisms, and syntactic structures—shapes the implementation of quantum logic. Current results suggest that expressiveness and productivity do not always align. Languages that offer rich abstractions or high-level constructs may introduce cognitive or lexical overhead, while those that minimize control structures and token usage can improve productivity without necessarily increasing expressive power. Moreover, compact functional construct present in languages like Qmod, appear to support concise and efficient algorithm design, suggesting a promising direction for future quantum language development. In conclusion, expressiveness and productivity in quantum programming are related but not equivalent. Each language exhibits trade-offs between abstraction, readability, and implementation effort. Future language development should aim to bridge the abstraction gap by enabling high-level quantum logic expression while maintaining manageable structural and cognitive complexity.

We also demonstrated that despite quantum programs bearing substantial differences contrasted against quantum ones, software complexity metrics in the quantum domain proved effective in capturing structural differences between programming languages. These metrics can serve as a foundation for more evaluations that integrate qualitative dimensions such as readability, maintainability, and learning curve. As a consequence, formally measurable attributes we associate with expressiveness do not translate into productivity in a one-to-one relationship. Evaluating dimensions of language performance from a cognitive standpoint \cite{green1989cognitive} is essential to understand how language design impacts the scalability, clarity, and feasibility of quantum software development.

In the future, we will pursue several research directions. The most immediate one is extending the list of quantum programming languages and algorithms to evaluate: we need to uncover how expressiveness scales with algorithmic complexity. In terms of metrics, the evaluation framework needs to incorporate additional software metrics, including maintainability indexes, readability scores, or memory usage to capture broader aspects of developer productivity and code quality. Qualitative methods such as developer surveys or usability studies can complement quantitative data to understand how programmers perceive and interact with different abstractions.

In response to the initial question of this study --\textit{Is productivity in quantum programming equivalent to expressiveness?}-- the results indicate that are distinct yet interrelated dimensions of a highly complex cognitive activity. Quantum programming languages may exhibit high expressiveness through rich syntax and abstractions but still require greater implementation effort. In contrast, having a compact and efficient syntax may enhance productivity without necessarily offering broad expressive capabilities. Measuring both aspects separately and in conjunction is essential to move the field forward.

\section*{Acknowledgements}
This work was supported by CENFOTEC University through the Excellence Scholarship Program in Quantum Computing. The authors would like to thank Ignacio Trejos-Zelaya for valuable guidance around programming language design and evaluation metrics. S.N-C. thanks the National Center for Supercomputing Applications and the Illinois Quantum Information Science and Technology Center for continued support, and  CENFOTEC University for the opportunity to establish a fruitful research collaboration.

\bibliographystyle{IEEEtran} 
\bibliography{references}

\vspace{12pt}

\end{document}